\documentclass[one column]{revtex4}
\usepackage[utf8]{inputenc}
\usepackage{hyperref}
\usepackage{pdflscape}

\usepackage{subfigure}
\usepackage{epsfig}
 \usepackage{epstopdf}
\usepackage{amsmath,amsfonts}

\begin{document}
\title{{ A modified thermodynamics of rotating and charged BTZ black hole } }
 \author{Sudhaker Upadhyay${}^{a,b,c}$}
 \email{sudhakerupadhyay@gmail.com; sudhaker@associates.iucaa.in}
 \author{Nadeem-ul-islam${}^{d}$}
  \email{drnadeemulislam@gmail.com}
\author{Prince A. Ganai${}^{d}$}
  \email{princeganai@nitsri.com}

 \affiliation{${}^{a}$Department of Physics, K.L.S college, Nawada, Bihar-805110, India} \affiliation{${}^{b}$Department of Physics, Magadh University, Bodh Gaya,
 Bihar  824234, India}
   \affiliation{${}^{c}$Visiting Associate, Inter-University Centre for Astronomy and Astrophysics (IUCAA) Pune, Maharashtra-411007, India}

 \affiliation{${}^{d}$Department of Physics, National Institute of Technology, Srinagar, Jammu \& Kashmir-190006, India }

\begin{abstract}
We present the thermodynamics of a charged and rotating BTZ black holes here. In  particular, we  derive
expressions for various macroscopic thermal quantities such as entropy, Hawking temperature,
Helmholtz free energy,  internal energy, enthalpy, Gibbs free energy, and specific heat.
To study the effects of  small statistical thermal fluctuations around the
equilibrium on thermodynamics, we calculated the leading-order corrections to the various
thermodynamical potentials of charged and
rotating BTZ black hole and do comparative analysis for the fixed values of charge and angular momentum.
\end{abstract}
\maketitle
\section{Overview and motivation}
 Long ago, it was proposed that a black hole behaves as a thermodynamic object. This  hypothesis was based on a logical analogy from
 the classical theory of general relativity where the square of irreducible mass \cite{q} (or in other words the area of event horizon \cite{w})
 never decreases. A similar kind of behavior is replicated in thermodynamics for entropy,  i.e., the entropy of any thermodynamical system never decreases in principle. This fundamental behavior of entropy (an increase of entropy) is quoted as
 the second law of thermodynamics. Thus by analogy, we have  the second law of black hole thermodynamics where the role of
 entropy is assumed to be played by area of the event horizon. This  analogy marked the beginning of a new subject
 that is black hole thermodynamics.
 For the zeroth and first law
of black hole thermodynamics, surface  gravity plays the role of the temperature of standard thermodynamics.
There exists one to one correspondence between the  classical thermodynamics and black hole thermodynamics \cite{mann}.

Later  on, the quantitative relation between the entropy and area of the event horizon  was established in 1973 \cite{e,e0}. It is
 found that entropy and
 area of event horizon are related through the relation $S = \frac{A}{4}$.
 It should be emphasized here that the role of pressure in  thermodynamics is played
 by the cosmological constant. Initially, people doubted that the second law of thermodynamics is getting violated due to the assumption that anything that falls into a black hole can never come out. Therefore, there is a loss of entropy (information). To rescue the second law of
 black hole  thermodynamics from being violated
these were assigned a  maximum entropy \cite{r,t,y}.
 However, the assignment of maximum entropy to black holes led
 to the discovery of holographic duality \cite{u,i}. This pictures the degrees of freedom in any region of space-time to
 its boundary surface and  the  area-law gets corrected when the size of the black hole is reduced  (due to small statistical thermal fluctuations around equilibrium).
 How does entropy undergo  modification by these thermal fluctuations  is a very serious question.

In order to resolve this question,  people tried different approaches but ended up  with the same conclusion that the leading-order corrections  to small-sized black holes are logarithmic.
 For instance,  Rademacher expansion of partition was done and resulted logarithmic corrections to the entropy
 of black hole \cite{o}. The same results were found in the case of string-black hole correspondence too
 \cite{p}.  The correction to entropy due to thermal fluctuations leads to modification in the various thermodynamical
 equations of states \cite{sud0,sud1,sud2,sud3,sud4,sud5,sud6,sud7}.
 The effect of small thermal
 fluctuations on the  BTZ black hole \cite{sud7}, G\"odel black hole \cite{a} and on the massive black
 hole in Ads space \cite{s,s111}  had been analyzed. Moreover, the  first-order leading corrections to
 the Schwarzschild Beltrami-de-sitter black hole \cite{d}  and  dilatonic black hole
 \cite{f} were emphasized.  The effect of thermal fluctuations around
 equilibrium of small-sized black hole was also studied using the partition function approach  \cite{g}.     The phase structure
 of the charged  Reissener-Nordstr\"om black hole was discussed in \cite{k}. It was observed that they show non-equilibrium
 second-order phase transition. In \cite{l}, Hawking and   Page showed that when the temperature of Ads Schwarzschild
 black hole reaches a particular value, called as critical temperature, it undergoes  a phase transition.
 This laid  foundation for the study of the P-V criticality of black holes. Later on, a
 lot of progress has been made  and Hawking's work has been extended to other complicated
 Ads space-times too \cite{z,x}. Such investigations led to a close analogy between the charged black hole and
 Van der walls liquid-gas system.
Further,   Faizal et al. used an adaptive  model of graphene to study some thermodynamic properties of a black hole   \cite{1}.
The effect of thermal fluctuation on the properties of BTZ  black hole in massive gravity has been investigated  \cite{11}.
In Ref. \cite{111},   Hawking radiation  is studied by   tunnelling formalism. Moreover, the effect
of thermal fluctuation on the thermodynamics of black hole geometry with a hyperscaling violation has been discussed in Ref. \cite{sud4}. These thermal
fluctuations arise from quantum corrections to geometry describing the system under consideration. Corrections to STU black hole have been
computed in \cite{v}. These corrections which arise from thermal fluctuations affect the stability of the STU black hole. The
 effect of leading order corrections to dumb holes (analogues to a black hole), as a consequence of thermal fluctuation,
 has been studied
\cite{v1}. The effect of small statistical thermal fluctuations on the entropy  of singly spinning Kerr-Ads black hole has been
investigated in \cite{v11}. Furthermore, the corrected thermodynamic
equations of states were computed for dilatonic black Saturn \cite{v111}.   From the perspective of logarithmic
corrections to thermodynamics of the modified Hayward  black hole,  it was found that for such black holes stability is not affected by quantum
fluctuations \cite{1x}.  Recently, emergence of quantum fluctuations to geometry  from thermal fluctuations  is proposed  \cite{xxx}.

  Here in this paper, we consider  the charged  and rotating BTZ black hole  as two different  thermal systems and
  discuss their thermodynamics, motivated from entropy-area law.
 Following a close analogy between the classical thermodynamics and black hole thermodynamics,
 we  first study various thermodynamical equations of states and potentials of the  rotating BTZ black holes.
 For instance, starting from Hawking temperature and entropy for a rotating BTZ metric function,  we calculate Helmholtz free energy,
 internal energy,
volume, pressure, enthalpy, and Gibbs free energy of the system in equilibrium.
In search of an answer to the question that what happens when small stable fluctuations around the equilibrium of the thermal
system are taken into
account, we calculate leading-order corrections to the entropy of the rotating BTZ black hole. In order to
study the effects of such corrections on the behavior of entropy, we  plot the entropy with respect to the event horizon
radius for different values of correction parameter and observe that the limiting entropy ($\alpha=0$) curve at the saddle point is
an increasing function and takes only positive values. As expected, the thermal fluctuations
affect the entropy of small-sized black holes. We  observe two critical values of entropy for black holes
where thermal fluctuations do not affect.
   Between these
  critical points, entropy has a positive (negative) peak for the positive (negative) correction parameter.
  Before, the first
  critical point, corresponding to positive values of correction parameter,  the micro-canonical
entropy takes a negative asymptotic value which is  physically meaningless and forbidden.
We also notice that the entropy corresponding to the negative correction parameter takes a positive asymptotic value.
Once the  Hawking temperature and corrected entropy are known, we derive various  corrected
thermodynamical potentials of the rotating BTZ
black holes
to study the effect
of thermal fluctuations. In this regard, we start by computing the leading-order corrected enthalpy energy of the system
and  observe that for small black holes, the enthalpy takes positive (negative) asymptotic value
corresponding to positive (negative) correction parameter. We also notice that there exists a
critical point.  Beyond this
critical value, enthalpy  increases with the horizon radius.  Furthermore, to estimate the possible amount of energy available
for doing work, we resort to the derivation of  Helmholtz free energy.
We observe three critical points for free energy. Two critical points occur in the positive region and one occurs in the negative region.
Before the first critical point, opposite to the positive correction parameter, the free energy with a negative correction parameter takes a negative asymptotic value. We then explore the effect of quantum fluctuations on the volume.
  Once, enthalpy, entropy, temperature, pressure, and volume become known, we estimate the leading
  order corrections to
  internal energy and Gibbs free energy.
 We then scrutinize the effect of thermal fluctuations on the stability of the black hole by studying the nature of corrected specific heat
 The specific heat becomes positively valued (Corresponding to the negative value of the correction parameter)
for small-sized black holes, implying the introduction of
 stability in the system. A positive value of the correction parameter results in the negative value of specific heat at
 small horizon radius.
 After this, we present a brief review of the thermodynamics of non-rotating charged BTZ black hole. The effect of thermal fluctuations on various equations of states of charged BTZ black hole is
 seen via the derivations of
  various thermodynamic variables following the same trend as that of uncharged and rotating one.
  The effect of quantum fluctuations, on the entropy, is similar
  to that as on the entropy of uncharged rotating BTZ black hole. For the free energy of charged BTZ black hole, we observe
    two critical points. The first one occurs on the horizon axis while the other one occurs
  in negative region.
Before the first critical point, opposite to the negative correction parameter, the free energy with a negative correction parameter,
takes
negative asymptotic value. After the first
critical point,  the correction parameter does not play a significant difference in the free energy. Later on,
    perturbed enthalpy for charged BTZ black hole is calculated, and it is found that for black holes of the small event horizon,
    the enthalpy takes negative (positive) asymptotic value for  the positive (negative) correction parameter.  We
  also observe the existence of a critical point at a small horizon radius. From the perturbed enthalpy, we derive perturbed volume.
  We then
  make the use of corrected enthalpy and volume with pressure as an independent variable, for the
  derivations of other thermodynamic potentials like internal energy and Gibbs free energy. From the perturbed Gibbs free energy,
  we observe a critical point  beyond which Gibbs free energy remains constant. Before, the critical point,
   the correction parameter of positive nature, decreases the Gibbs free energy asymptotically. On the contrary, negative valued
   correction
   parameter  asymptotically increases the Gibbs free energy.  Corrected internal energy is then studied and it is found that
   the
a positive value of the correction parameter does not affect the internal energy much and  make the least difference by following the
same trend as that of
uncorrected internal energy curve. On the other hand, the negative value of the correction parameter yields a negative asymptotic value
of internal energy.
However as the size of the black hole  goes on increasing, the difference between the perturbed and unperturbed internal energy gets
minimized. Finally, we investigated the stability of charged BTZ black hole and we observed that
 thermal fluctuations, affect the specific heat in the same fashion as that for rotating one.  \\

   This work is presented as follows.
 In Sec. \ref{sec1}, we discuss the thermodynamics of  rotating BTZ black hole and derive the expressions for various
 equations of states. Within this section, we study  the effect of thermal fluctuations on the entropy of
   rotating BTZ black hole and determine   various leading-order corrected thermodynamic variables.
 The stability of   rotating BTZ holes under the effect of thermal corrections is  also studied.
 In Sec. \ref{sec5}, we briefly present the thermodynamics of   charged but stationary BTZ black holes and
 discuss the thermal instability in various thermodynamical variables.  We also study
  the stability  of charged BTZ black hole.
  Finally, in the last section \ref{sec8}, we summarize our results under the heading final remarks.

\section{Thermodynamics of  rotating BTZ black hole}\label{sec1}
In this section, we discuss the thermodynamics of the uncharged rotating BTZ black hole.
Let us start by writing a metric for the  rotating BTZ black hole as follows,
\begin{equation}
 ds^2 = -f(r)dt^2 +\frac{ dr^2}{f(r)} +  r^2(d{\varphi} + g({\varphi})dt^2)^2,
 \end{equation}
 where the expressions of metric functions are
\begin{equation}\label{m}
 f(r) = -8GM  + \frac{r^2}{l^2} +\frac{ 16 G^2 J^2}{r^2},
\end{equation}
and
\begin{equation}
 g({\varphi}) = -  \frac{4GJ}{r^2}.
\end{equation}
From the given lapse function $f(r)$, Eq.(\ref{m}), it is easy to calculate Hawking temperature  $T_H$ as
follows,
\begin{equation} \label{te2}
T_H = \left.\frac{f^\prime(r)}{4\pi}\right|_{r=r_+} = \frac{r_+}{2{\pi}l^2} - \frac{8G^2J^2}
{{\pi}r_+^3},
\end{equation}
where $r_+$ is the  horizon radius of black hole obtained by $f(r) =0$.
Using the fundamental postulate of black hole thermodynamics, the corresponding Bekenstein
entropy is calculated by
\begin{equation}\label{s02}
S_0 = \frac{{\pi}r_+}{2G}.
\end{equation}
This is the same as that of a stationary BTZ black hole. This is because of the fact that
the angular moment does not affect the area of the black hole as such entropy of BTZ black hole is not
modified by endowing rotations  to it.

Now, for a given entropy  Eq. (\ref{s02}) and Hawking temperature Eq. (\ref{te2}), we can derive the expression for Helmholtz free energy
(denoted by $F_r$, subscript $r$, is used to indicate the free energy of rotating BTZ black hole) of  rotating BTZ black hole as following:
\begin{eqnarray}
 F_r &= &-\int{S_0dT_H},\nonumber\\
 & =& -\int dr_+\left(\frac{\pi{r_+}}{2G}\right)\left( \frac{1}{2\pi{l^2}} +\frac{24{G^2}{J^2}}{\pi{r_+^4}}\right).
\end{eqnarray}

Upon solving the above integral, we get
\begin{equation}
 F_r = -\frac{{r_+}^2}{8Gl^2} + \frac{6GJ^2}{{r_+}^2} .
\end{equation}
This is the required expression for the Helmholtz free energy of  rotating BTZ black hole.
Moreover, it is a matter of calculation to  compute the enthalpy energy ($H_r$) of the system
using the following standard formula:
\begin{equation}
 H_r  = \int{T_HdS_0}.
\end{equation}
 Plugging the corresponding values of temperature  (\ref{te2}) and entropy   (\ref{s02}) in the above expression and
 we have
\begin{equation}
  H_r = \frac{{r_+}^2}{8l^2G} + \frac{2GJ^2}{{r_+}^2}.
\end{equation}
This expression refers to  enthalpy energy of  rotating BTZ black hole.

The expression for pressure is given by
\begin{equation}\label{1}
 P = \frac{1}{8 \pi Gl^2}.
\end{equation}
Writing $H_r$, in terms of $P$, we obtain,
\begin{equation}
  H_r = { P \pi{r_+}^2} + \frac{2GJ^2}{{r_+}^2}.
\end{equation}
 Hence,
the thermodynamic volume is calculated as
\begin{equation}
 V_r= \frac{d H_r}{dP}= { \pi{r_+}^2}.
\end{equation}

By using the above-derived quantities, we are in state of
calculating the further properties of the system such as internal energy ($U_r$) and Gibbs free energy ($G_r$). In fact,
it is well-known  that internal ($U_r$) is
 given mathematically  by following relation:
\begin{equation}\label{h}
U_r = H_r - PV_r.
\end{equation}

Substituting the corresponding values of enthalpy energy, pressure  and volume in Eq. (\ref{h}),  we have
\begin{equation}
 U_r = \frac{2GJ^2}{{r_+}^2} .
\end{equation}
This is the expression for internal energy of  rotating BTZ black hole.
In the same fashion, we would like to derive the expression for Gibbs free energy($G_r$). The thermodynamical formula
for Gibbs free energy is given by,
\begin{equation}
 G_r = F_r + PV_r.
\end{equation}
For a given values of    $F_r$,      $P$   and $V_r$,
the Gibbs free energy is calculated by
\begin{equation}
 G_r = \frac{6J^2G}{{r_+}^2}.
\end{equation}

In the forthcoming sections, we would like to see the consequences of  small stable fluctuations
around equilibrium  on the thermodynamics of rotating BTZ black holes.

\subsection{The leading-order perturbed entropy}\label{sec2}
To calculate the entropy of a thermal system, there are two main ways: (a) by considering a microcanonical ensemble, where
an equilibrium system is treated as being genuinely isolated; (b)
by considering a canonical ensemble, where the system is treated as being in thermal
contact with a very large reservoir at a fixed temperature.
For the strongly self-gravitating
black holes at equilibrium the canonical
ensemble consideration cannot be a good choice.
For our case, we follow the latter case as the canonical consideration works well for the small-sized thermal system.
We calculate the entropy of  rotating  BTZ black hole when small stable fluctuations
around equilibrium are taken into account. For the canonical black holes,  the most convenient starting point is the partition function
      \begin{equation} \label{e}
       Z(\beta) = { \displaystyle \int_{0}^{\infty}{{dE}\rho{(E)}e^{-\beta{E}}}},
      \end{equation}
      where       $\beta$  is the inverse of  hawking temperature  (for convenience Boltzmann constant is set to unit). Here,
 density of states $ \rho{(E)}$ can be calculated easily for a given  partition function by
 taking  inverse Laplace transform of Eq. \ref{e}
       as follows
\begin{eqnarray}
       \rho{(E)} &=&\frac{1}{2\pi{i}}{ \displaystyle\int_{{\beta}_0 - i\infty}^{{\beta}_0+ i\infty}{{d\beta}Z(\beta)e^{\beta{E}}}},\nonumber\\
& =& \frac{1}{2\pi{i}}{ \displaystyle\int_{{\beta}_0 - i\infty}^{{\beta}_0+ i\infty}{{d\beta}e^{\ln{Z(\beta)} + \beta{E}}}}.\label{l}.
\end{eqnarray}
Here one should note that the exponential term is nothing but the exact entropy of a   black hole, i.e.
      \begin{equation}
    {\cal S}(\beta) = \ln{Z(\beta)} + \beta{E}.
      \end{equation}
    This exact entropy is the total sum of entropies of all individual subsystems and depends on temperature explicitly.
      The role of thermal fluctuations is very significant only if the size of the black hole
       is a very small  i.e   event horizon radius is very small and this also justifies
       the canonical consideration of the ensemble.
        To study the effects of thermal fluctuations on entropy we expand  ${\cal S}(\beta)$  around equilibrium,
       via Taylor expansion as following:
      \begin{eqnarray} \label{uet}
      {\cal S}(\beta) = S_0 + \frac{1}{2}{({\beta} -  {\beta}_0})^2  \left.
      \frac{d^2{\cal S}}{d\beta^2}\right|_{{\beta}= {\beta}_0}   + \mbox{(higher-order   terms)},
      \end{eqnarray}
            where  $S_0$ indicates the canonical entropy at saddle point equilibrium  (temperature).
  Plugging this expanded value of entropy from Eq. \ref{uet}  in Eq. \ref{l},   we  have the following density of states:
            \begin{equation}
       \rho{(E)} = \frac{e^{S_0}}{2\pi{i}}{\int{{d\beta}e^{\frac{1}{2}{({\beta} -
       {\beta}_0})^2{\frac{d^2{\cal S}}{d{\beta}^2}}}}}.
      \end{equation}
  The further simplification of this integral leads to the density of states
      \begin{equation}
       \rho{(E)} = \frac{e^{S_0}}{\sqrt{2\pi{\frac{d^2{\cal S}}{d{\beta}^2}}}}.
      \end{equation}
    The logarithm of density of states gives the microcanonical entropy at equilibrium
      \begin{equation}
      S = \ln{\rho} = S_0 - \frac{1}{2}\ln{\frac{d^2S}{d{\beta}^2}} +
      \mbox{ (sub leading  terms)}.
     \end{equation}
 This  relation  holds for   all thermodynamic systems (including black
holes) considered as a canonical ensemble.
   The correction term is estimated by \cite{das}
     \begin{equation}
      \frac{d^2{\cal S}}{d{\beta}^2} = C{T^2_H},
     \end{equation}
   where $C$ represents  the dimensionless specific heat.
   This leads to the following expression for the microcanonical entropy at equilibrium
      \begin{equation}
      S = \ln{\rho} = S_0 - \frac{1}{2}\ln{C{T^2_H}} +
      \mbox{ (sub-leading  terms)}.
     \end{equation}
     In case of BTZ black holes the specific heat coincides with the  equilibrium   value  of
     canonical entropy  at saddle point $S_0$
     as there are no work terms. This suggest the leading-order entropy  at equilibrium
     \begin{equation}
      S = S_0 -\frac{1}{2}\ln{S_0{T^2_H}}.
     \end{equation}
    Without loss of generality, we replace coefficient $\frac{1}{2}$ in second term by a  more general correction   parameter ``$\alpha$". This leads to the most general  form of corrected entropy
     \begin{equation}\label{n}
      S = S_0 - \alpha{\ln{S_0{T^2_H}}}.
     \end{equation}
     Here   ${\alpha}$ characterizes  the effect of thermal  fluctuations on the entropy of system. We observe here that the leading-order
 correction  to canonical entropy of black hole is  logarithmic  in nature.
For our considered system of BTZ black holes, the canonical entropy at saddle point is known
from Bekenstein-Hawking area law, from Eq. (\ref{s02}). Substituting
the values of Hawking temperature, from (\ref{te2}) and  Bekenstein-Hawking area law  entropy from   (\ref{s02})  in
Eq. (\ref{n}),  we get
the micro-canonical entropy of rotating BTZ black hole  at equilibrium as follows,
\begin{equation}\label{cs}
S_c = \frac{{\pi}r_+}{2G} - \alpha \ln{\frac{({r_+}^4 - 16G^2 J^2 l^2)^2}{8\pi Gl^2 {r_+}^5}}.
\end{equation}
To have a comparative analysis between the corrected and uncorrected entropy  at equilibrium, we plot the obtained expression for corrected
entropy  against the event horizon radius for different values of correction parameter.

\begin{figure}[htb]
\includegraphics[width=80 mm]{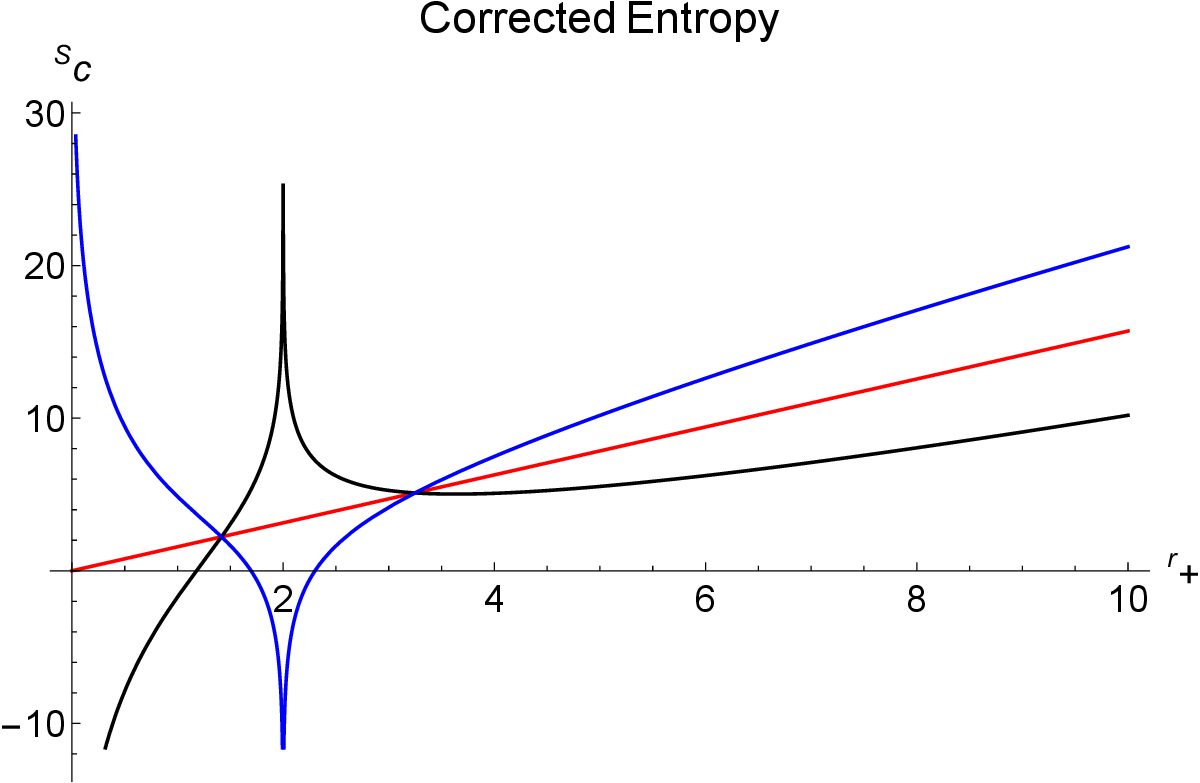}
\caption{Entropy  vs. the black hole horizon. Here $\alpha=0$ denoted by red line, $\alpha=1.5$ denoted by black
curve,  $\alpha=-1.5$ denoted by blue curve.}
 \label{fig1}
\end{figure}
From the plot (FIG. \ref{fig1}), we observe that in the limit $\alpha \rightarrow {0}$, the  original entropy curve  at saddle  point is shown in red color which is an increasing function and
takes positive values only.
However,  entropy shows surprising results when quantum corrections are taken
into account
 for small-sized black holes. However, we notice that the behavior of entropy for large black holes does
 not undergo any deformation as anticipated.
This implies that quantum corrections to entropy are significant only for black holes of very small horizon radius. Also,
there exist two critical points. The entropy in between these critical points shows a positive peak for the positive value of the correction
parameter, while for negative values of the correction parameter, the corrected entropy shows a negative peak in between the
critical points. The negative values of entropy are physically meaningless and therefore forbidden. However,  for small black holes
whose horizon radii are smaller than the first critical horizon radius, the behavior of corrected entropy is
reversed,
i.e corresponding to positive values of correction parameter the micro-canonical entropy leads to the negative asymptotic value which
is again physically meaningless. The micro-canonical entropy with a negative correction parameter takes a positive asymptotic
value and therefore leads to the extra stability in the system. Beyond the second critical point, the micro-canonical entropy
in tandem to area-law entropy is an  increasing function of event horizon radius irrespective of different specific
values of the correction parameter. This justifies the insignificance of thermal fluctuations at a larger event horizon radius.
 \subsection{The leading-order corrections to thermodynamic potentials}\label{sec3}
\begin{figure}[htb]
\includegraphics[width=80 mm]{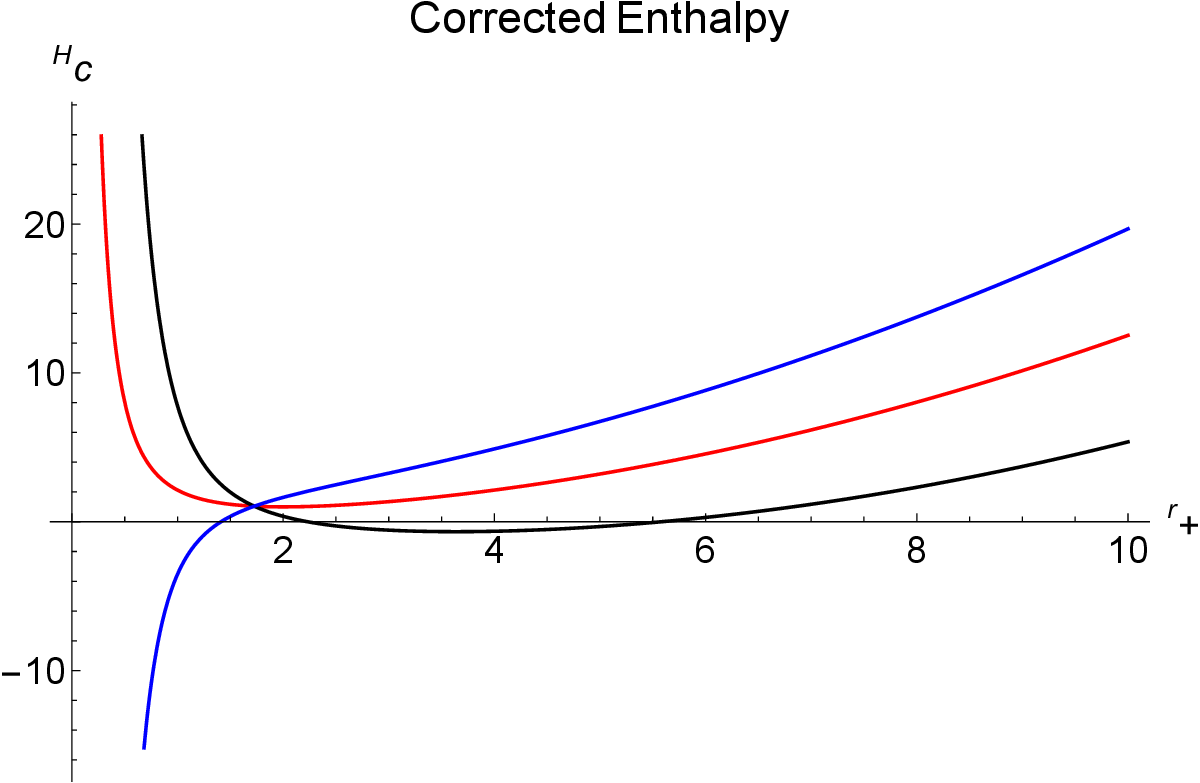}
\caption{Enthalpy  vs. the black hole horizon. Here $\alpha=0$ denoted by red line, $\alpha=1.5$ denoted by black
curve,  $\alpha=-1.5$ denoted by blue curve.}
 \label{fig2}
\end{figure}

 In this section, we would like to evaluate various thermodynamic variables to define the state of the system (ensemble). In this regard, we start with the enthalpy of the
 system.
 It is mathematically defined  as follows
 \begin{equation}\label{ent}
  H_c = \int{T_HdS_c}.
 \end{equation}
 where $H_c$ represents corrected enthalpy energy.
Plugging the resulting values for Hawking temperature from Eq.  (\ref{te2}) and  corrected Bekenstein entropy
from Eq. (\ref{cs}) in Eq. (\ref{ent}), we obtain
\begin{equation} \label{in}
 H_c = \frac{\alpha 40 G^2 J^2 }{3 \pi {r_+}^3} + \frac{2 G J^2}{{r_+}^2} -\frac{3 \alpha { r_+}}{2 \pi l^2} + \frac{{r_+}^2}{8 G l^2}.
\end{equation}
This is  a leading-order corrected expression for Enthalpy  considering thermal fluctuations in account.
This leading-order correction terms reflect
 the  effect of thermal fluctuations around the equilibrium.
To describe  the effect of quantum fluctuations analytically, we plot the
resulting enthalpy energy against the event horizon radius for fixed value of angular moment in FIG. \ref{fig2}.
 Once again from the plot (FIG.\ref{fig2}), it is quite clear   that thermal fluctuations are effective only at small event
 horizon radius while
 the enthalpy of large sized black holes is unaffected.
For black holes of the small event horizon, the enthalpy takes a negative asymptotic
  value for  the negative correction parameter, however, it takes a positive asymptotic value for the positive correction parameter.  We
  also notice that there exists a critical point beyond which internal energy is increasing function of event horizon radius.

\begin{figure}[htb]
\includegraphics[width=80 mm]{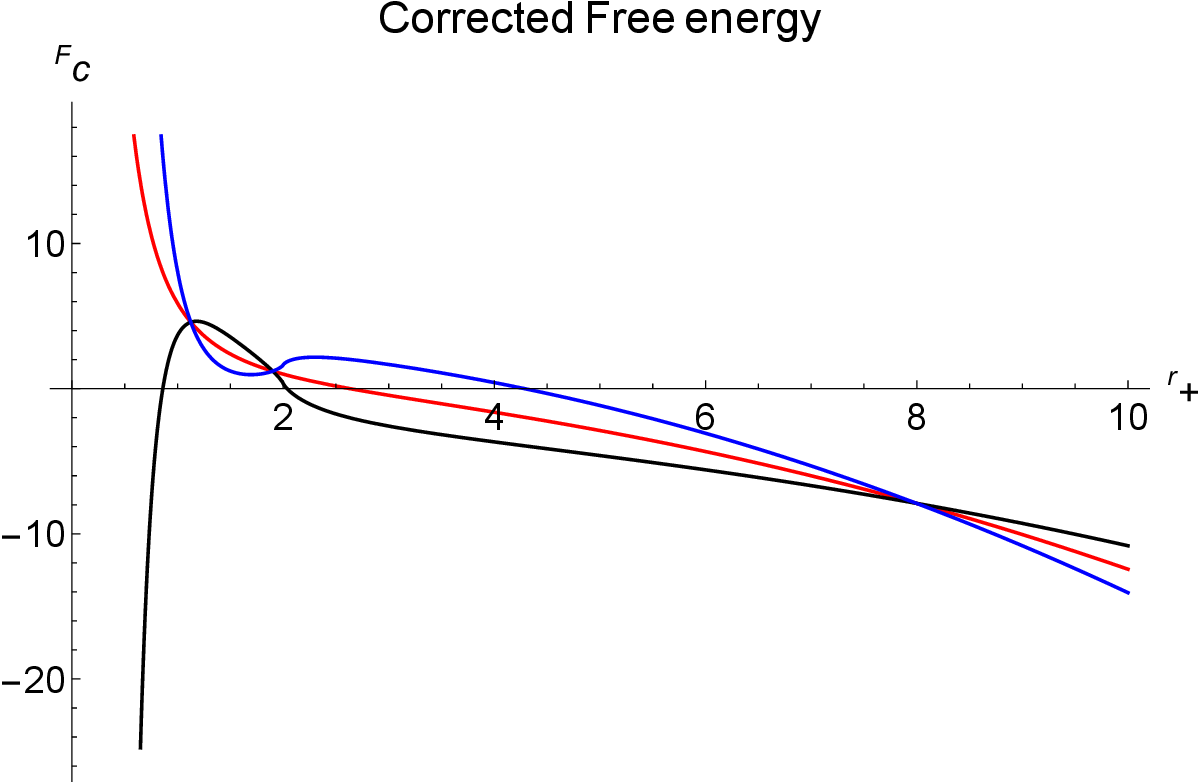}
\caption{Free energy  vs. the black hole horizon. Here $\alpha=0$ denoted by red line, $\alpha=1.5$ denoted by black
curve,  $\alpha=-1.5$ denoted by blue curve.}
 \label{fig3}
\end{figure}

Besides enthalpy energy, there are three more important thermodynamic potentials to describe
the state of the system, namely Helmholtz free energy, internal energy, and Gibbs free energy.
We analyze the effects of thermal fluctuations on these thermodynamic potentials one by one. Being a state function, free energy
   represents the possible amount of energy available for doing work.
    Quantitatively, Helmholtz free energy is
  defined  as follows
  \begin{equation}
   F_c = - \int{S_cdT_H}.
  \end{equation}
  Here $F_c$ denotes corrected free energy.
 On plugging the   values of leading-order corrected Bekenstein entropy and Hawking temperature,  the above definition leads
 \begin{eqnarray}\label{hel}
F_c  &=&\frac{\alpha r_+ \log \left(\frac{\left({r_+}^4-16 G^2 J^2 l^2\right)^2}{8 \pi  G l^4 {r_+}^5}\right)}{2 \pi  l^2}-
\frac{8 \alpha G^2 J^2 \log \left(\frac{\left(r^4-16 G^2 J^2 l^2\right)^2}{8 \pi  G l^4 r^5}\right)}{\pi  r^3}\nonumber\\
&+&
\frac{40 \alpha G^2 J^2}{3 \pi  {r_+}^3}-\frac{3 \alpha r_+}{2 \pi  l^2}+\frac{6 G J^2}{{r_+}^2}-\frac{{r_+}^2}{8 G l^2}.
\end{eqnarray}
To analyze the effect of thermal fluctuations on the free energy,  we plot the obtained expression (\ref{hel}) with respect to the event horizon radius in FIG.  \ref{fig3}.
From the plot, we observe three critical points for free energy. Two critical points occur in positive region and one occurs in negative region.
Before the first critical point, opposite to the positive correction parameter, the free energy with negative correction parameter
takes negative asymptotic value. After the first
critical point,  the correction parameter does not play a significant difference for the free energy.

Once we know the corrected expression of  corrected enthalpy
 energy, it is a matter of calculation to  find the leading-order corrected  volume as following:
 \begin{equation}
 V_c=  \frac{d H_c}{dP}\big|_{j,r} = \pi{r_+}^2  -12 \alpha G r_+.
\end{equation}
\begin{figure}[htb]
\includegraphics[width=80 mm]{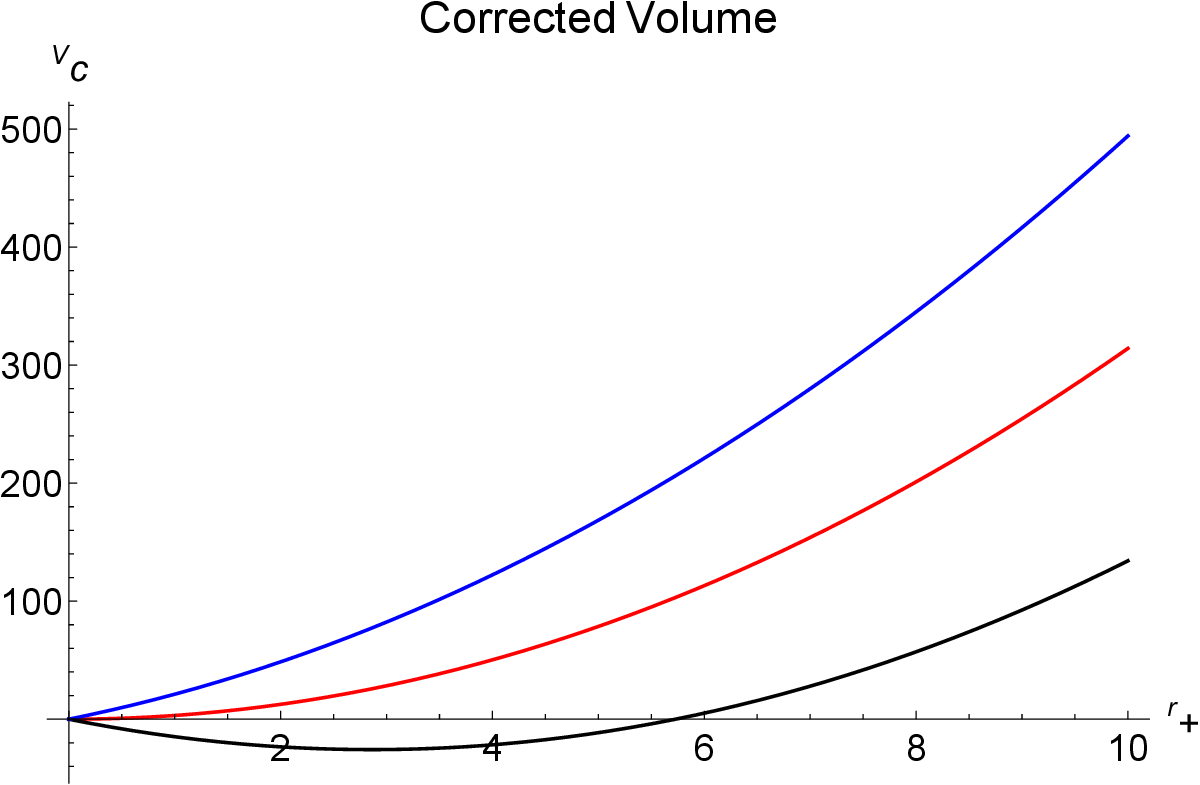}
\caption{Volume  vs. the black hole horizon. Here $\alpha=0$ denoted by red line, $\alpha=1.5$ denoted by black
curve,  $\alpha=-1.5$ denoted by blue curve. }
 \label{fig4}
\end{figure}

  Furthermore,  we derive the corrected expression for internal energy. Internal energy being a state function has profound significance in
  thermodynamics and represents the sum total  of energy of
 basic building blocks of a thermodynamic system, i.e., the sum of kinetic energy arising from the motion of particles and potential energy
 by virtue of the particular configuration of these particles.
  The corrected internal energy ($U_c$) can be calculated from the following formula:
 \begin{equation}
  U_c = H_c - PV_c.
 \end{equation}
   Now we have at hand both corrected enthalpy energy and corrected volume.
   With these quantities we derive the leading order corrections
   to internal energy as fallows,
 \begin{equation} \label{two}
 U_c = \frac{\alpha 40 G^2 J^2 }{3 \pi {r_+}^3} + \frac{2 G J^2}{{r_+}^2}.
 \end{equation}
 This represents a quantitative measure of the effect of quantum fluctuations on internal energy. To draw a parallel, between
uncorrected  and corrected internal energy, the Eq. \ref{two}, is plotted against the horizon radius. From the plot, we observe that
for small-sized black holes, the uncorrected internal energy is large, however, it goes on decreasing as the size of the black hole increases. The
positive value of the correction parameter  does not affect the internal energy much and  makes the least difference by following the same
trend as that of
uncorrected internal energy curve. On the other hand, the negative value of the correction parameter yields a negative asymptotic value of internal energy.
However as the size of the black hole goes on increasing, the difference between the uncorrected and corrected internal energy gets minimized. It is obvious from the above expression and given plot that in the limit $\alpha$ tends to zero,
 the uncorrected internal energy is retrieved,
   as expected.

 \begin{figure}[htb]
\includegraphics[width=80 mm]{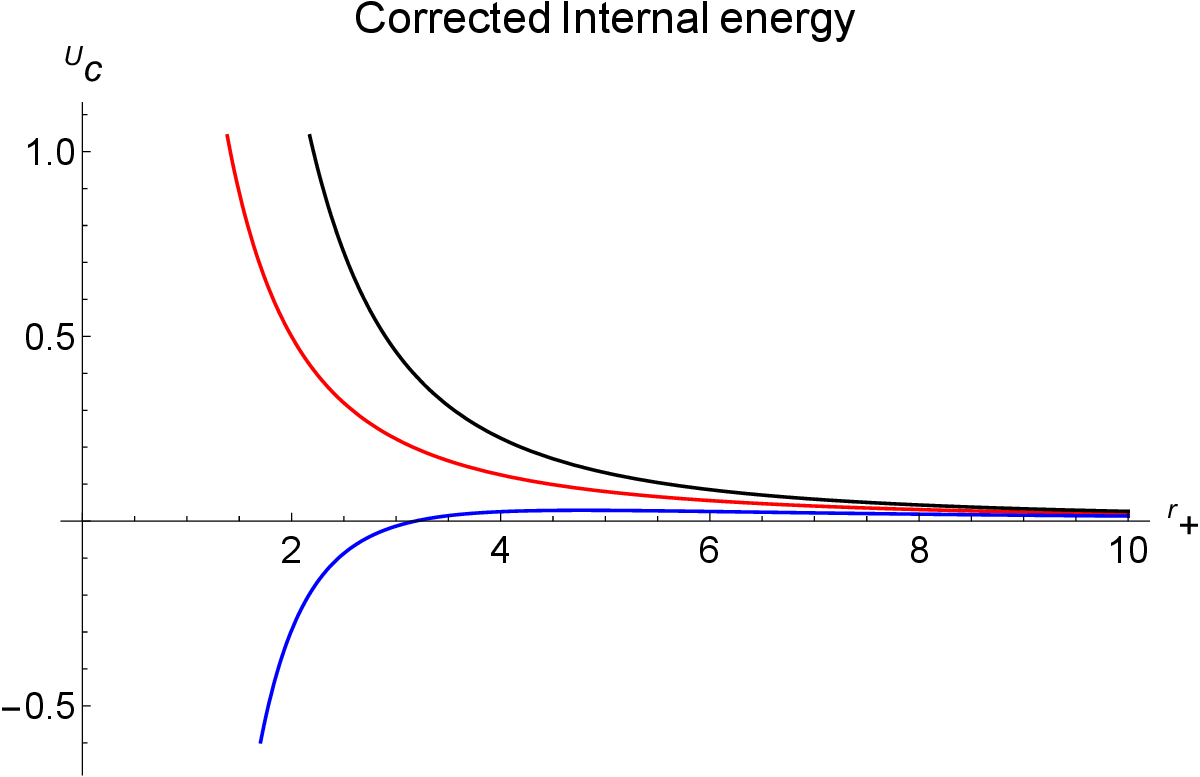}
\caption{Enthalpy  vs. the black hole horizon. Here $\alpha=0$ denoted by red line, $\alpha=1.5$ denoted by black
curve,  $\alpha=-1.5$ denoted by blue curve.}
 \label{fig5}
\end{figure}

   We  will now explore the effects of thermal fluctuations on the fourth thermodynamic potential i.e Gibbs free energy.
   Gibbs free energy in thermodynamics measures  the maximum amount of mechanical work that can be extracted from a system. It
   is mathematically  represented by the relation given below,
   \begin{equation}
     G_c = F_c + PV_c,
   \end{equation}
   where the symbols have their usual meanings
   Inserting the  corrected values of pressure and free energy we obtain,
  \begin{eqnarray}\label{gib}
G_c =\frac{80 \alpha G^2 J^2 l^2+3 \alpha \left({r_+}^4-16 G^2 J^2 l^2\right) \log{\frac{\left({r_+}^4-
16 G^2 J^2 l^2\right)^2}{8 \pi  G l^4 {r_+}^5}}-18 \alpha {r_+}^4+36 \pi  G J^2 l^2 r_+}{6 \pi  l^2 {r_+}^3}.
\end{eqnarray}
\begin{figure}[htb]
\includegraphics[width=80 mm]{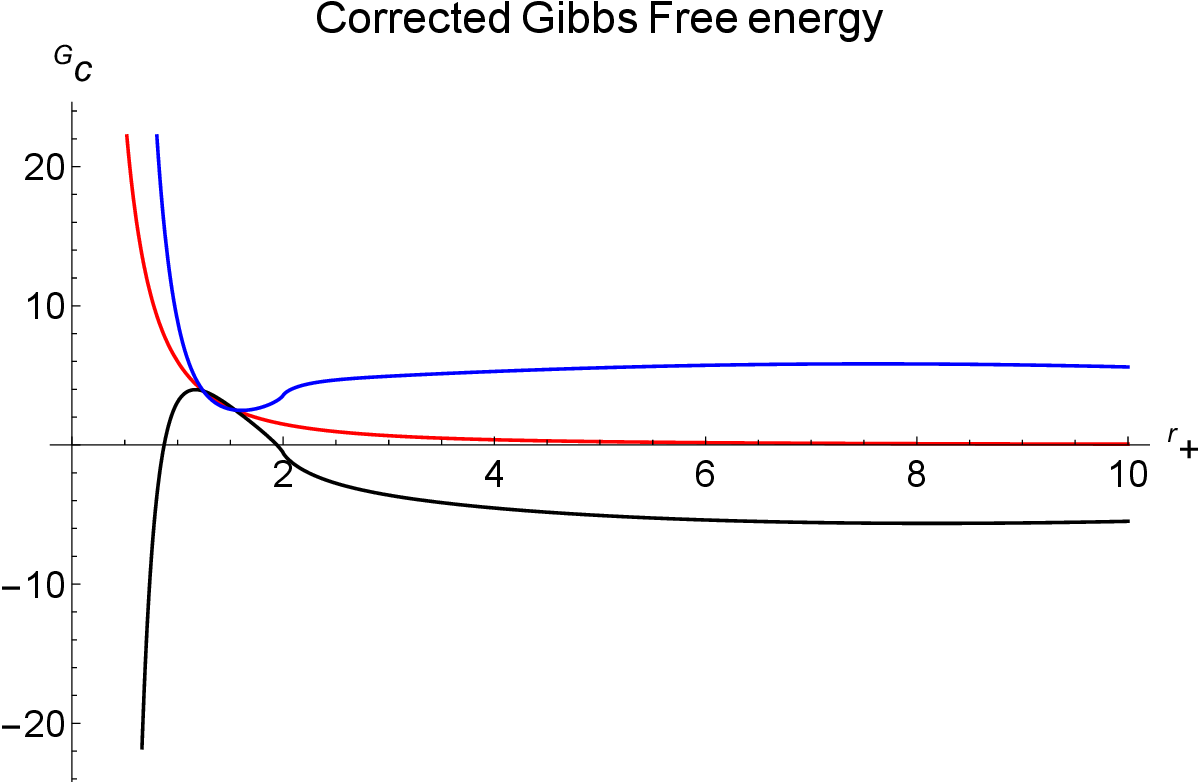}
\caption{Gibbs free energy   vs. the black hole horizon. Here $\alpha=0$ denoted by red line, $\alpha=1.5$ denoted by black
curve,  $\alpha=-1.5$ denoted by blue curve.}
 \label{fig6}
\end{figure}\\

We thus have corrected expression for Gibbs free energy. To appreciate  the difference between the corrected and
uncorrected Gibbs free energy,
 we plot the obtained expression, i.e, Eq. \ref{gib} against the event horizon radius.

From the plot, we observe that Gibbs free energy undergoes a significant change in its behavior as $r_+$  tends to zero.
We found two critical points at a small event horizon radius.
Between these two critical points, thermal fluctuations compel Gibbs free energy to undergo a minuscule but important change.
In this region, the positive value of
the correction parameter leads to a slight increase  in Gibbs free energy, while the negative correction parameter
performs the opposite operation in the same region.  Before the first critical point,
one can examine that  the positive (negative) value of the correction parameter produce a negative (positive) asymptotic value of
Gibbs free energy. Negativity in Gibbs free energy  shows the sign of stability in this region and  hints at the maximum amount of
energy that can
 be extracted from the system for useful work
\subsection{Stability of rotating BTZ black hole}\label{sec4}
  In order to explore the stability of black holes we study the nature of its specific heat. From the nature of specific heat,  we can estimate whether black hole
  undergoes phase transition or not.
  The positive values of specific heat confirm that the system is stable against the phase transition,   while the negative value of specific heat infers the instability of system.
  We  derive the expression for  specific heat by taking thermal fluctuation in account
    which must reduce to the original expression for uncorrected specific heat when
    fluctuation is switched-off (i.e. $\alpha$ equal to zero). The specific heat ($C_c$) can be
  derived with the help of following  well-known formula from classical thermodynamics:
      \begin{equation}
       C_c = T_H\frac{d{S_c}}{dT_H}.\label{sp}
      \end{equation}
  Inserting the  values of the corrected internal energy  from Eq. (\ref{in}) and Hawking temperature
  from Eq.  (\ref{te2}) in the  above definition, i.e Eq. (\ref{sp}),  we can easily have the expression
       for leading-order corrected specific heat for the black holes. This reads as
\begin{equation}\label{spe}
C_c =-\frac{160 \alpha G^3 J^2 l^2+6 \alpha G {r_+}^4+16 \pi  G^2 J^2 l^2 r_+-\pi  {r_+}^5}{96 G^3 J^2 l^2+2 G {r_+}^4}.
\end{equation}
To analyze the effect of small statistical  fluctuations around equilibrium on the stability of our system we plot the
corrected specific heat (Eq. \ref{spe}) against the event horizon radius for fixed values of  angular moment.
\begin{figure}[htb]
\includegraphics[width=80 mm]{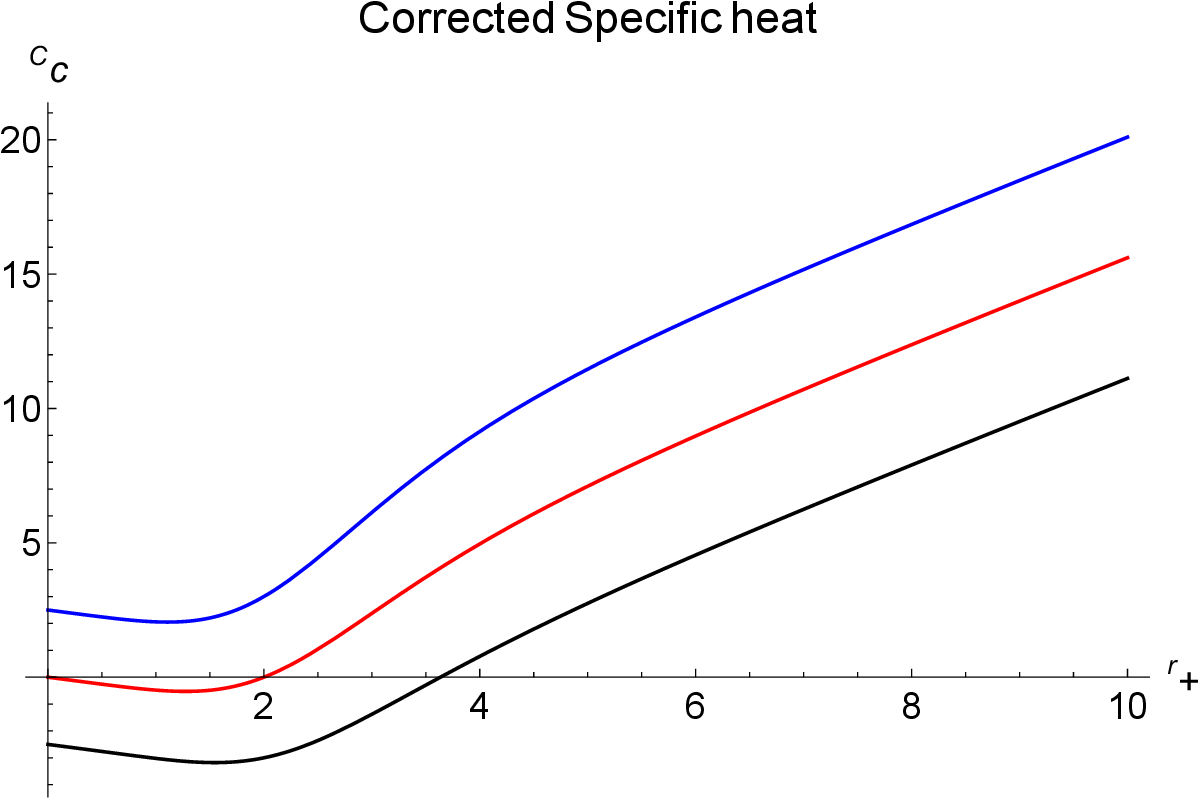}
\caption{Specific heat  vs. the black hole horizon. Here $\alpha=0$ denoted by red line, $\alpha=1.5$ denoted by black
curve,  $\alpha=-1.5$ denoted by blue curve. }
 \label{fig7}
\end{figure}
From the  plot (FIG. \ref{fig7}), we observe that if there are no thermal fluctuations in
consideration (i.e. in the limit $\alpha$ tends to zero), the specific heat becomes
 less  negative-valued for small black hole and takes positive value  for large sized black holes.
 This indicates that without thermal fluctuation, the small sized black holes
is unstable and large sized black holes are stable.  When thermal fluctuations around the equilibrium
are taken into consideration, the specific heat becomes positive valued (Corresponding to negative value of the correction parameter)
for small sized black holes also. It implies that thermal fluctuations of such a kind  brings
 stability in the system. Positive value of the correction parameter results in the negative value of specific heat at small horizon radius
 thereby enhancing the unstable behavior of black hole. For large black holes,  thermal fluctuations do not play any significant
 role  which is not surprising.

 \section{Thermodynamics of  charged BTZ black hole}\label{sec5}
 In this section, we first write the metric function characterizing the charged, but non-rotating BTZ black hole  as follows
\begin{equation}\label{mn}
 f(r) = -8GM  + \frac{r^2}{l^2} - \frac{\pi Q^2} {2}\ln{\frac{r}{l}}.
\end{equation}
Following the previous section,  the Hawking temperature is calculated as follows,
\begin{equation}\label{temi}
T_H = \left.\frac{f^\prime(r)}{4\pi}\right|_{r=r_+} = \frac{r_+}{2{\pi}l^2}  - \frac{Q^2}{8r_+},
\end{equation}
where $r_+$ is the  horizon radius of black hole obtained by $f(r) =0$.

The equilibrium value of entropy is same as that of uncharged one i.e,
\begin{equation}\label{s01}
S_0 = \frac{{\pi}r_+}{2G}.
\end{equation}
From the given Hawking temperature and equilibrium entropy, we obtain the free energy for charged BTZ black hole,
given by the following expression:
\begin{equation}\label{s012}
F = -\frac{{r_+^2}}{8G l^2}-
\frac{\pi  Q^2 \log{\frac{r_+}{l}}}{16 G}.
\end{equation}
The expression for enthalpy is derived, using the following well-known formula
\begin{equation}
 H = \int{T_HdS_0}.
\end{equation}
 Plugging the corresponding values of temperature  from Eq. (\ref{temi}) and entropy  from Eq. (\ref{s01}) in the above
 expression and
 by solving this, we have
\begin{equation}
  H = \frac{{r_+}^2}{8l^2G}  - \frac{Q^2 \pi}{16 G}\ln{\frac{r_+}{l}}.
\end{equation}  We can express for enthalpy   in terms of pressure as below,
\begin{equation}\label{ret}
  H = P \pi {r_+}^2  - \frac{Q^2{\pi}}{32G}\ln{ 8 \pi G  P{r_+}^2}.
\end{equation} From the Eq. (\ref{ret}), we obtain the expression for thermodynamic volume
(conjugate of pressure in thermodynamics) as follows
\begin{equation}
 V=  \frac{d H}{dP}\big|_{j,r} = \pi{r_+}^2  -\frac{{\pi} Q^2}{32 GP}.
\end{equation}
Having expressions for volume, Enthalpy and pressure, we  are able to derive internal energy as
\begin{equation}
 U =  \frac{{\pi} Q^2}{32G}[1- \ln{ 8 \pi G  P{r_+}^2}].
\end{equation}
The Gibbs free energy is defined by
\begin{equation}
G = F+ PV.
\end{equation}
Using the corresponding values of free energy, volume and pressure, one gets
\begin{equation}
 G = - \frac{Q^2\pi}{32G}(1 +2\ln\frac{r_+}{l}).
\end{equation}
This is the equilibrium value of Gibbs free energy for the charged BTZ black hole.
\subsection{The corrected thermodynamics of charged  BTZ black hole}\label{sec6}
Thermal fluctuations in thermodynamics mimic the quantum fluctuations in  space time geometry. As the black hole evaporates its size decreases
in other words it becomes hot as hawking temperature is inversely proportional to mass. Now hot black hole signifies the inevitability of
thermal fluctuation and hence quantum fluctuations. These fluctuations modify the thermodynamic behavior of system. Let us look at the corrected version of
entropy. To obtain this, we just substitute Eqs.   (\ref{temi})  and (\ref{s01}) in Eq. (\ref{n}), and     get  the following   perturbed entropy of charged BTZ black hole:
 \begin{equation}\label{1234}
{S_p}=\frac{\pi  r}{2 G}-\alpha \log{\frac{\pi  r \left(\frac{Q^2}{8 r}-\frac{r}{2 \pi  l^2}\right)^2}{2 G}}.
\end{equation}
 To have a comparative analysis between the perturbed and exact entropy  at equilibrium, we plot the expression for perturbed
entropy against the event horizon radius for different values of the correction parameter.

\begin{figure}[htb]
\includegraphics[width=80 mm]{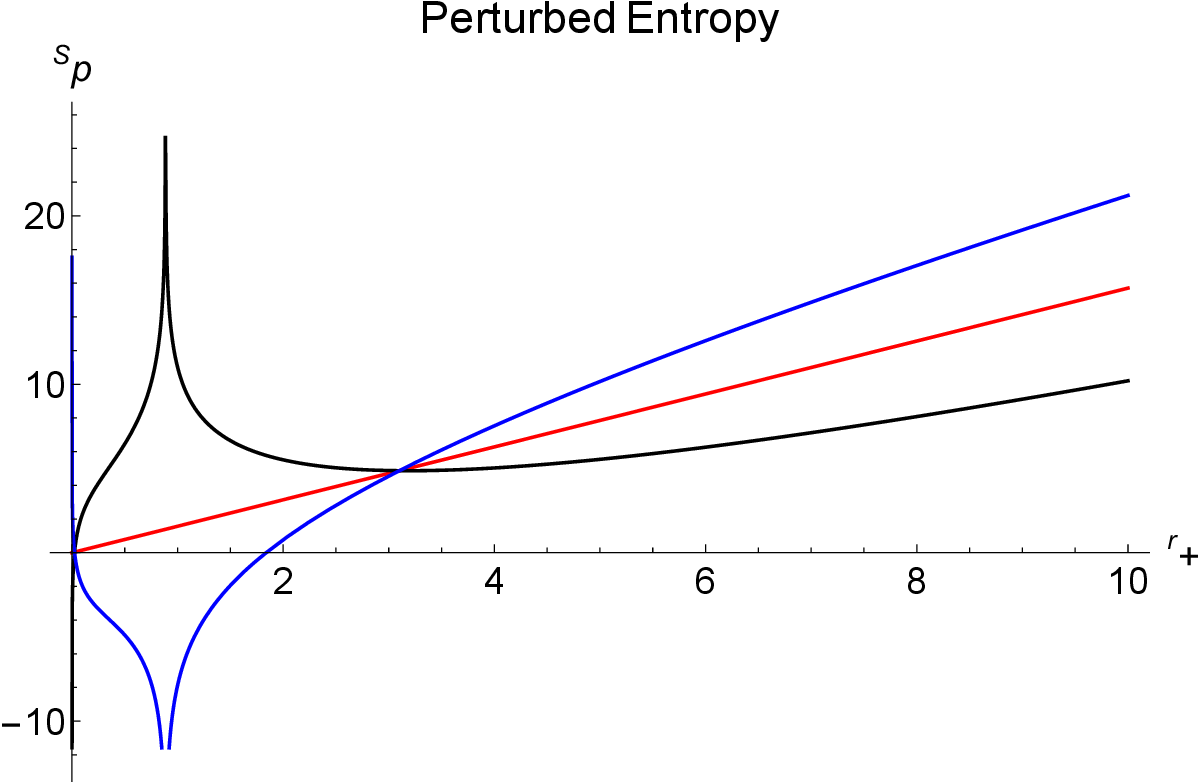}
\caption{Entropy  vs. the black hole horizon. Here $\alpha=0$ denoted by red line, $\alpha=-1.5$ denoted by blue
line,  $\alpha=-1.5$ denoted by blue line,  and
$\alpha=1.5$ denoted by green
line.}
 \label{fig11}
\end{figure}
From the plot (FIG. \ref{fig11}), it is observed that in the limit $\alpha \rightarrow {0}$  the  original entropy curve (as
shown in red color)  is an increasing function and
takes positive values only,  as expected.
However,  entropy shows surprising results when quantum corrections are taken
into account
 for small-sized black holes, in tandem to our assumption while deriving the  Generalized form of perturbed entropy
i.e Eq. (\ref{n}). We notice that the behavior of entropy for large black holes does
 not undergo any deformation as anticipated.
This implies that quantum corrections to entropy are significant only for black holes of very small horizon radius. Also,
there exist two critical points. The entropy in between these critical points shows a positive peak for the negative value of
the correction
parameter, while for negative values of the correction parameter, the corrected entropy shows a negative peak in between the
critical points. The negative values of entropy are physically meaningless and therefore forbidden.
 The first critical point happens to be close to the singularity, and before  the first critical point,
 the behavior of thermal fluctuations reverse
i.e corresponding to  positive values of correction parameter, the micro-canonical entropy leads to the negative asymptotic value which
is again physically meaningless. The micro-canonical entropy with  negative correction parameter takes a positive asymptotic
value and therefore leads to the extra stability in the system. Beyond the second critical point, the micro-canonical entropy
in tandem to area-law entropy is an  increasing function of event horizon radius irrespective of different specific
values of correction parameter. This justifies the insignificance of thermal fluctuations at a larger event horizon radius.

After observing the effect of quantum fluctuations on the entropy, we move on to reckon the perturbed free energy, following the same trend as that for
rotating BTZ black hole in the above section, and the derived expression reads,
\begin{eqnarray}\label{hel1}
F_p = \frac{1}{8} \left(\frac{\alpha \left(4 {r_+}^2-\pi  l^2 Q^2\right) \log{\frac{\pi  r_+ \left(\frac{Q^2}{8 r_+}-
\frac{r_+}{2 \pi  l^2}\right)^2}{2 G}}}{\pi  l^2 r_+}-\frac{12 \alpha r_+}{\pi  l^2}+\frac{\alpha Q^2}{r_+}-\frac{{r_+^2}}{G l^2}-
\frac{\pi  Q^2 \log{r_+}}{2 G}\right).
\end{eqnarray}
\begin{figure}[htb]
\includegraphics[width=80 mm]{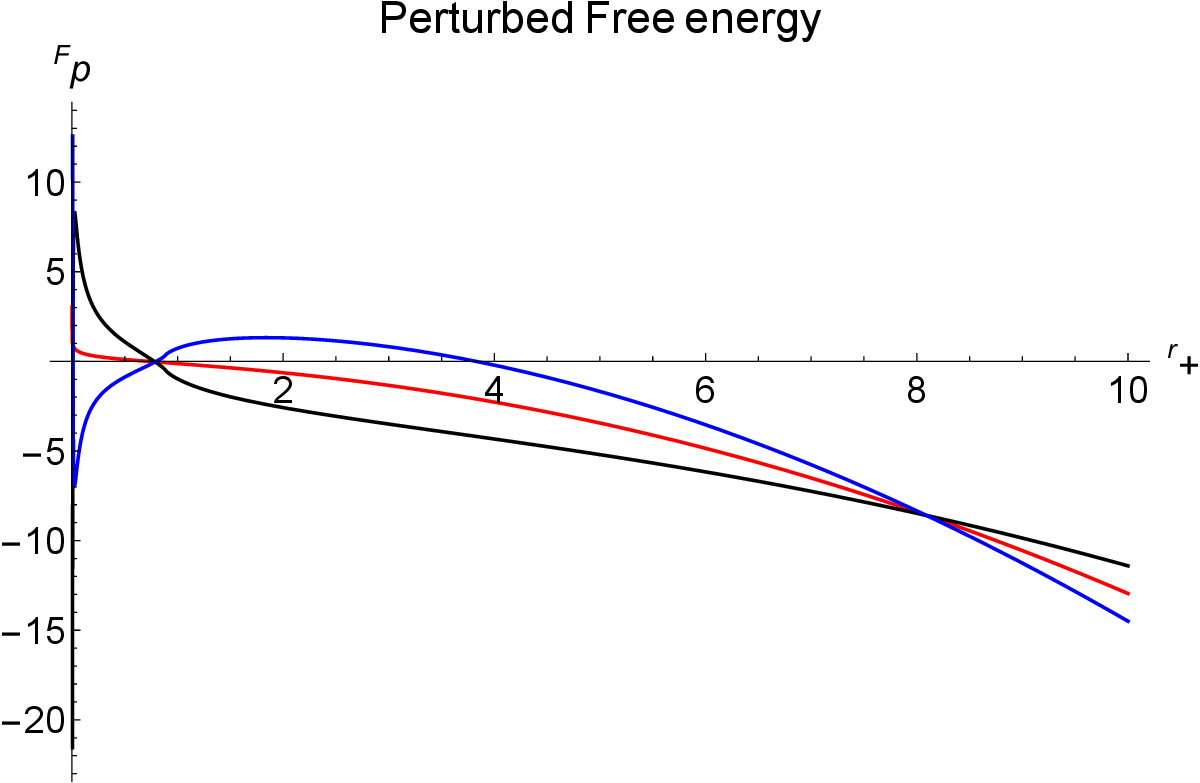}
\caption{Free energy  vs. the black hole horizon. Here $\alpha=0$ denoted by red line, $\alpha=1.5$ denoted by black
curve,  $\alpha=-1.5$ denoted by blue curve.}
 \label{fig333}
\end{figure}
To analyze the effect of thermal fluctuations on the free energy,  we plot the obtained expression (\ref{hel1}) with respect to
the event horizon radius in FIG.  \ref{fig333}.

\begin{figure}[htb]
\includegraphics[width=80 mm]{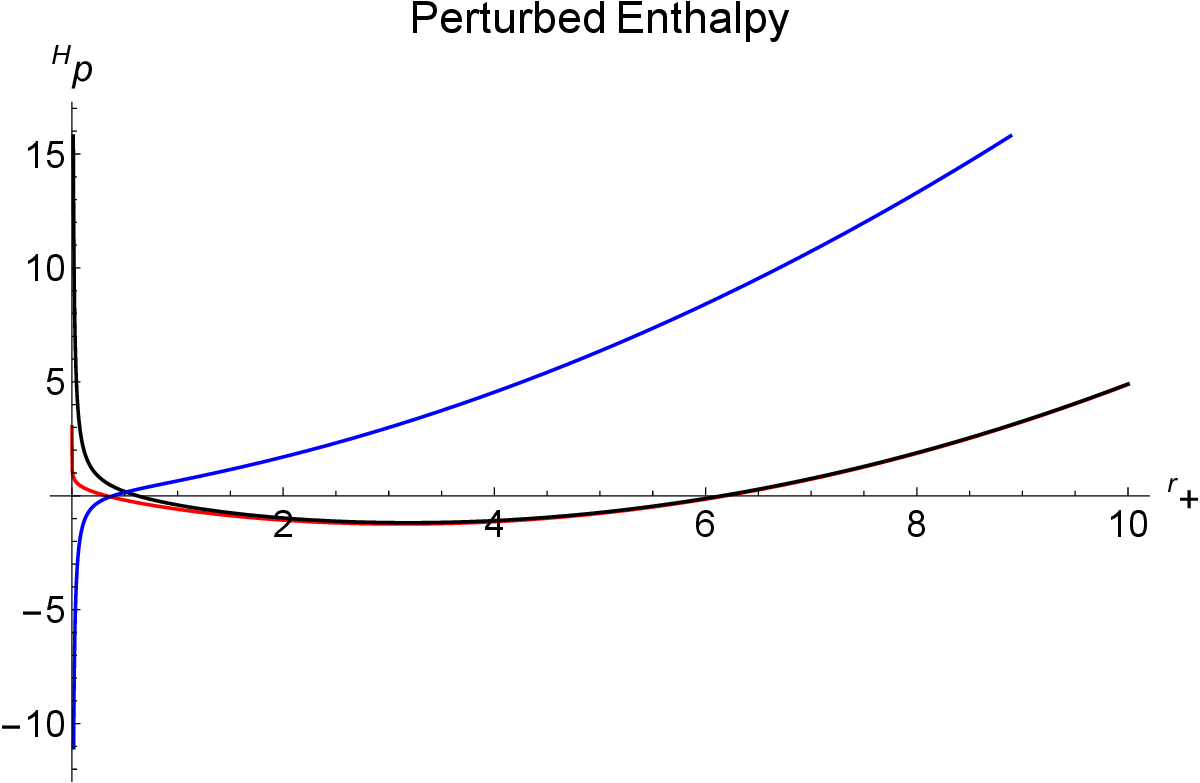}
\caption{Enthalpy  vs. the black hole horizon. Here $\alpha=0$ denoted by red line, $\alpha=1.5$ denoted by black
curve,  $\alpha=-1.5$ denoted by blue curve.}
 \label{fig331}
\end{figure}

From the plot (FIG. \ref{fig333}), we observe two critical points for free energy. The first one occurs on the horizon axis while as the other one occurs in negative region.
Before the first critical point, opposite to the negative correction parameter, the free energy with positive correction parameter takes
positive asymptotic value. After the first
critical point,  the correction parameter does not play a significant difference for the free energy.\\
 Once  the entropy and  Hawking temperature is known, it is a matter of computation to reckon the perturbed enthalpy. The quantitative
 expression for perturbed enthalpy is given by,
\begin{equation}\label{in1}
 H_p =\frac{1}{16} \left(-\frac{24 \alpha r_+}{\pi  l^2}+\frac{2 \alpha Q^2}{r_+}+\frac{2 {r_+}^2}{G l^2}
 -\frac{\pi  Q^2 \log{\frac{r_+}{l}}}{G}\right).
\end{equation}
In terms of pressure,
 \begin{equation}\label{int123}
 H_p =-12 \alpha G P r_+ + \frac{\alpha Q^2}{8 r}-\frac{\pi  Q^2 \log{8 \pi  G P r^2}}{32 G}+\pi  P {r_+}^2.
\end{equation}
In order to investigate the effect of quantum fluctuations analytically, we plot the
resulting enthalpy energy against the event horizon radius for a fixed value of charge in,  FIG. \ref{fig331}.

 Once again from the plot (FIG.\ref{fig331}), it is quite clear   that thermal fluctuations are effective only at a small event
 horizon radius while
 the enthalpy of large-sized black holes is unaffected.
For black holes of small event horizon, the enthalpy takes negative asymptotic
  value for  the negative correction parameter, however it takes positive asymptotic value
  for positive correction parameter.  We
  also notice that there exists a critical point beyond which perturbed enthalpy energy is increasing function of event horizon radius for
  negative value of correction parameter. While as it takes negative value from the critical point upto $r_+ = 6$, and then onwards increses with
   $r_+$, for positive values of correction parameter.
 From the perturbed enthalpy, one can find the expression for perturbed volume simply by differentiating the enthalpy w.r.t to pressure at
 fixed charge,
 \begin{figure}[htb]
\includegraphics[width=80 mm]{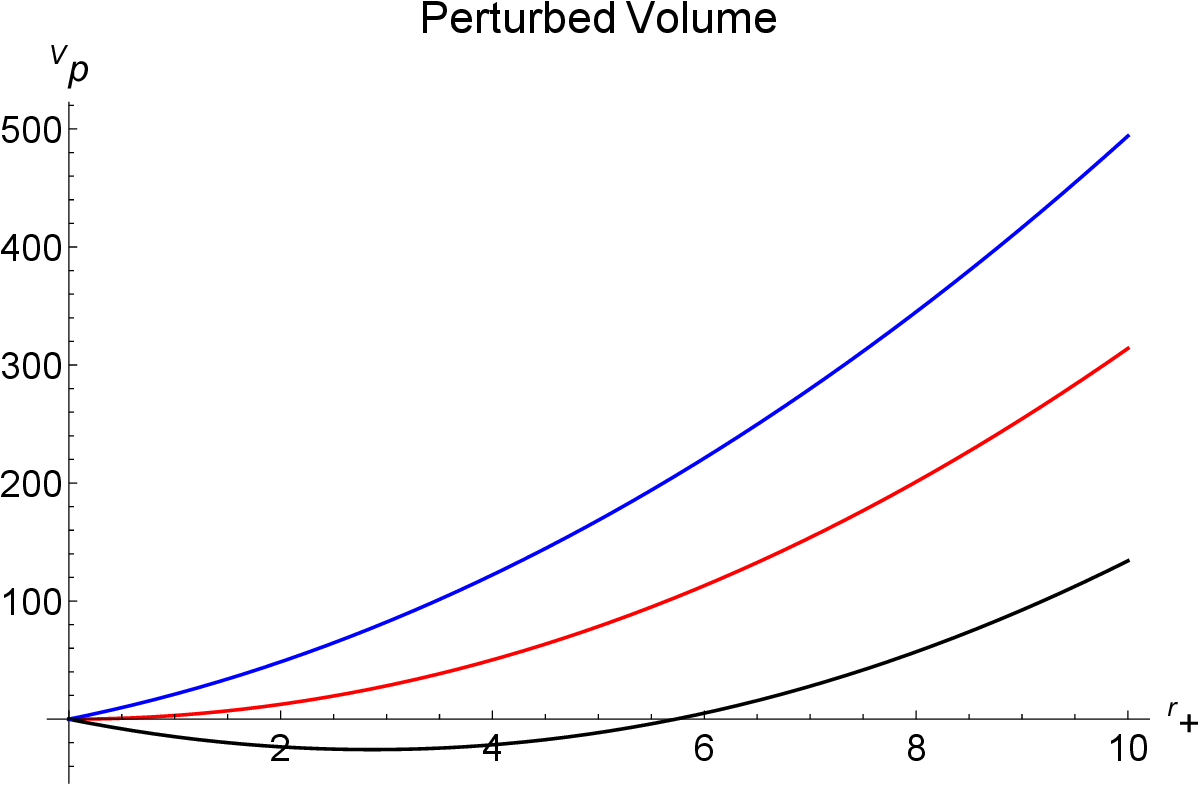}
\caption{Volume  vs. the black hole horizon. Here $\alpha=0$ denoted by red line, $\alpha=1.5$ denoted by black
curve,  $\alpha=-1.5$ denoted by blue curve.}
 \label{fig512}
\end{figure}
 \begin{equation}
 V_p=  \frac{d H_p}{dP}\big|_{j,r} = \pi{r_+}^2  -\frac{{\pi} Q^2}{32 P} - 12 \alpha G r_+.
\end{equation}
The effect of volume is described analytically in plot , FIG. \ref{fig512}. From the same plot, we found that
thermal fluctuations measured by negative
valued correction parameter, increase the volume and follows the same trend as that of original volume curve,
while as the positive valued correction parameter
initially decreases the volume up to certain limit, and produces a negative volume region, (which is physically forbidden) and then increases at
large distance implying the uselessness of quantum fluctuations at large
horizon radius

We now calculate the corrections to Gibbs free energy which reads as follows
\begin{eqnarray}
G_p= -\frac{Q^2\pi}{32G}(1 +2\ln{r_+}) +\frac{\alpha Q^2}{8r_+} +
 \frac{\alpha  \left(4 {r_+}^2+-\pi  l^2 Q^2\right) \log{\frac{\pi  r_+ \left(\frac{Q^2}{8 r_+}-
\frac{r_+}{2 \pi  l^2}\right)^2}{2 G}}}{8\pi  l^2 r_+} .
\end{eqnarray}
\begin{figure}[htb]
\includegraphics[width=80 mm]{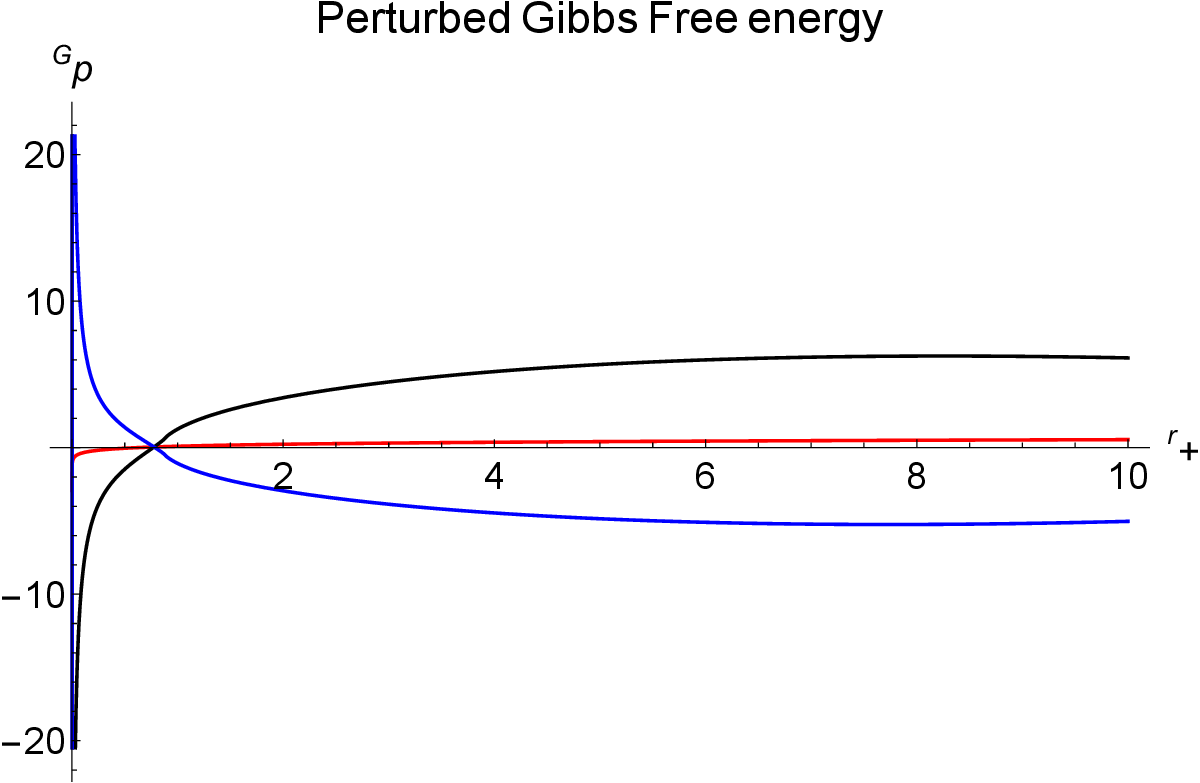}
\caption{Gibbs free energy   vs. the black hole horizon. Here $\alpha=0$ denoted by red line, $\alpha=1.5$ denoted by black
curve,  $\alpha=-1.5$ denoted by blue curve.}
 \label{fig61}
\end{figure}\\
 To analyze the effect of quantum fluctuation and hence thermal fluctuations qualitatively, we resort to graphical line of attack and plot
 the perturbed Gibbs free energy against the event horizon radius for fixed values of charge and different values of correction parameter
  as shown in  FIG. \ref{fig61}.
   From the plot  (FIG.\ref{fig61}), we, discern a critical point beyond which Gibbs free energy remains constant. Before, the critical point,
   the correction parameter of positive nature, decreases the Gibbs free energy asymptotically. On the contrary, the negative valued correction
   parameter leads to an asymptotic increase in Gibbs free energy before the critical point.\\

Another very important thermodynamic potential characterizing the BTZ black hole is internal energy and
the perturbed internal energy is given by
\begin{equation}\label{tui}
  U_p =    \frac{Q^2 \pi}{32 G}(1-2\log{\frac{r_+}{l}})  +\frac{\alpha Q^2}{8 r_+}.
\end{equation}

   It is obvious from the above expression that in the limit $\alpha$ tends to zero, we recover the uncorrected internal energy
   as expected.
 In order to have a qualitative analysis of these corrections, we plot the corrected internal energy against the event horizon radius
 for
  various values of correction parameter and observe the consequent effects on the internal energy of the considered system.
 \begin{figure}[htb]
\includegraphics[width=80 mm]{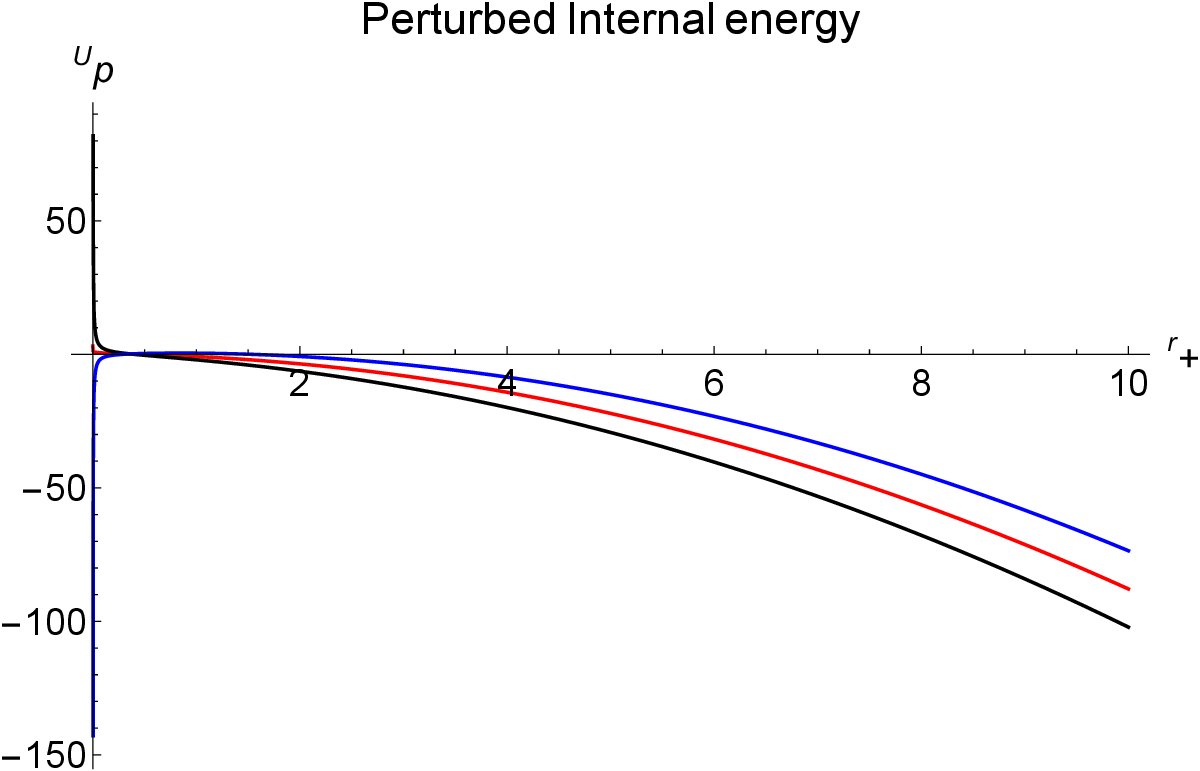}
\caption{Enthalpy  vs. the black hole horizon. Here $\alpha=0$ denoted by red line, $\alpha=1.5$ denoted by black
curve,  $\alpha=-1.5$ denoted by blue curve.}
 \label{fig500}
\end{figure}
We   notice that at larger event horizon   radius the thermal fluctuations fail to affect
our system. From the plot, we observe a critical point, beyond the critical point,   enthalpy is an decreasing function
of event horizon radius irrespective of nature of correction parameter.
However,  before the  critical point, the nature of the correction
parameter is of profound importance. In this region, the negative value of the correction parameter $\alpha$ produces a
negative asymptotic decrease. On the other hand, the positive value of correction parameter  produces quite opposite behavior
and hence increases the internal energy considerably.\\

 \subsection{Stability of  charged stationary BTZ black hole}\label{sec7}
 In order to explore the stability of black holes, once again, we resort to stability parameter called specific heat. The specific heat is given by
      \begin{equation}\label{sp1}
       C_c = T_H \frac{dS_p}{dT_H}.
      \end{equation}
  Inserting the  values of the perturbed entropy  from Eq. (\ref{1234}) and Hawking temperature from Eq.  (\ref{temi}) in the  above
definition (\ref{sp1}),  we can easily have the expression
       for leading-order perturbed specific heat for the black holes. This reads as
\begin{equation}\label{spe2}
C_p=-\frac{2 \pi  \alpha G l^2 Q^2+24 \alpha G {r_+}^2+\pi ^2 l^2 Q^2 r_+-4 \pi  {r_+}^3}{2 \pi  G l^2 Q^2+8 G {r_+}^2}.
\end{equation}
To analyze the effect of small statistical  fluctuations around equilibrium on the stability of our system we plot the
corrected specific heat (\ref{spe2}) against the event horizon radius for fixed values of electric charge
\begin{figure}[htb]
\includegraphics[width=80 mm]{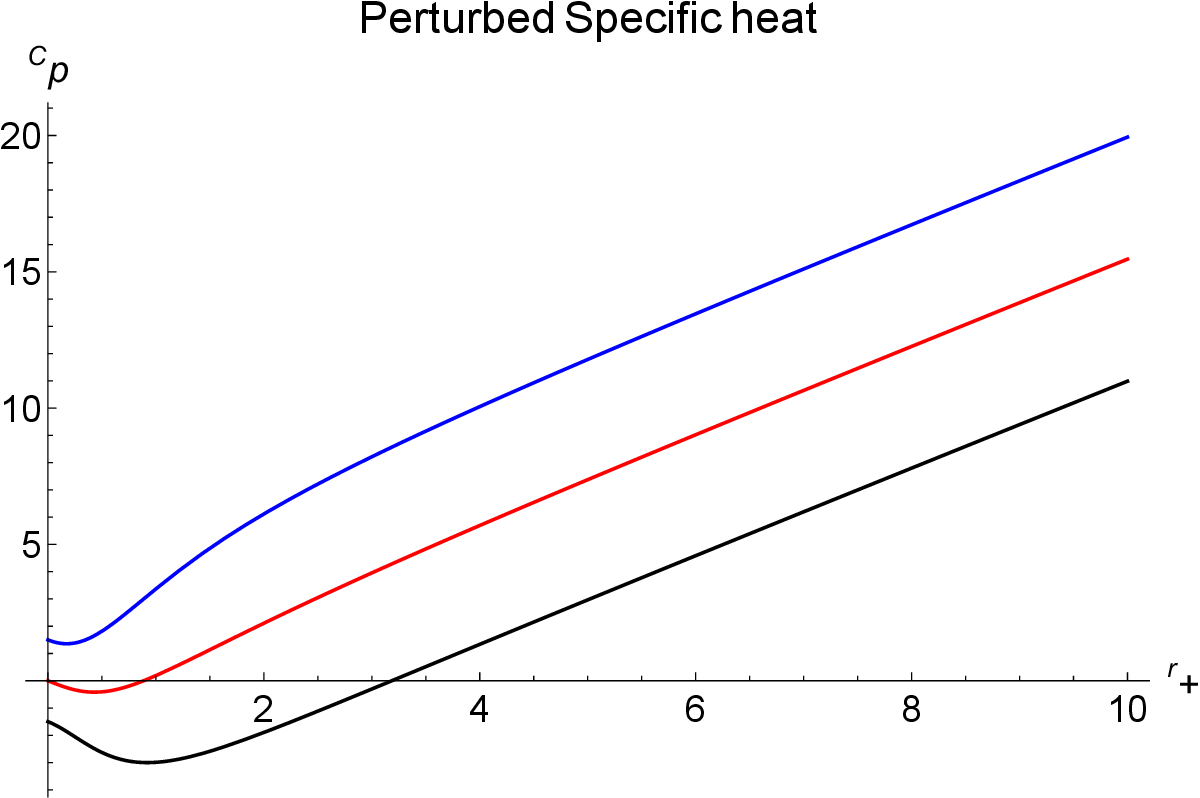}
\caption{Specific heat  vs. the black hole horizon. Here $\alpha=0$ denoted by red line, $\alpha=1.5$ denoted by black
curve,  $\alpha=-1.5$ denoted by blue curve.}
 \label{fig70}
\end{figure}
From the  plot (FIG. \ref{fig70}), we found that absence of thermal fluctuations in charged black holes
 (i.e. in the limit $\alpha$ tends to zero), the specific heat becomes
 less  negative-valued for small black hole and takes positive value  for large sized black holes. The perturbed specific heat
 capacity for charged BTZ  follows the similar trend as that of rotating but uncharged BTZ black hole.
  When thermal fluctuations around the equilibrium
are taken into consideration, the specific heat becomes positive valued (Corresponding to negative value of the correction parameter)
for small sized black holes. It implies that thermal fluctuations of such a kind  brings
 stability in the system in tandem to that of rotating BTZ black hole. Positive value of the correction parameter results
 in the negative value of specific heat at small horizon radius
 thereby producing the instability in such  black holes.\\
\section{Final remarks}\label{sec8}
  Here we have considered  the charged  and rotating BTZ black holes  as two different  thermal systems and
  discussed their thermodynamics, motivated from entropy-area law.
 Following a close analogy between the classical thermodynamics and black hole thermodynamics,
 we first studied various thermodynamical equations of states and potentials of the  rotating BTZ black holes.
 For instance, starting from Hawking temperature and entropy for a rotating BTZ metric function,  we calculated Helmholtz free energy,
 internal energy,
volume, pressure, enthalpy, and Gibbs free energy of the system in equilibrium.
In search of an answer to the question that what happens when small stable fluctuations around the equilibrium of the thermal
system are taken into
account, we calculated leading-order corrections to the entropy of the rotating BTZ black hole. In order to study the effects of such corrections on the behavior of entropy, we have plotted entropy with respect to the event horizon
radius for different values of correction parameter and have found that the limiting entropy ($\alpha=0$) curve at the saddle point is
an increasing function which takes only positive values. As expected, the thermal fluctuations
affect the entropy of small-sized black holes. We have found two critical values of entropy for black holes
where thermal fluctuations do not affect.
   Between these
  critical points, entropy has a positive (negative) peak for the positive (negative) correction parameter.
   The negative values of entropy are physically meaningless and so forbidden.
  Before, the first
  critical point, corresponding to positive values of correction parameter,  the micro-canonical
entropy takes a negative asymptotic value which is again physically meaningless and forbidden.
We have also found that entropy corresponding to the negative correction parameter, takes a positive asymptotic value which,
therefore, gives extra stability to the system.

Once the  Hawking temperature and corrected entropy became known, we derived various thermodynamical potentials of the rotating BTZ
black holes
to study the effect
of thermal fluctuations. In this regard, we have started by computing the leading-order corrected enthalpy energy of the system
and have found that for small black holes the enthalpy energy takes positive (negative) asymptotic value
corresponding to positive (negative) correction parameter. We also noticed that there exists a
critical point, where the effects of thermal fluctuation are irrelevant.  Beyond this
critical value, enthalpy energy increases with the horizon radius.  Furthermore, to estimate the possible amount of energy available
for doing work, we have derived Helmholtz free energy.
We observed three critical points in free energy. Two critical points are found in the positive region and one occurred in the negative region.
Before the first critical point, opposite to the positive correction parameter, the free energy with negative correction parameter
takes negative asymptotic value  We then explored the effect of quantum fluctuations on the volume.
  Once, enthalpy, entropy, temperature, pressure, and volume became known, we estimated the leading order corrections to
  internal energy and Gibbs free energy.
 We then tried to understand the effect of thermal fluctuations on the stability of the black hole by studying the nature of corrected specific heat
 The specific heat becomes positively valued (Corresponding to negative value of the correction parameter)
for small-sized black holes, implying the introduction of
 stability in the system. A positive value of the correction parameter results in the negative value of specific heat at
 small horizon radius.
 After this, we present a brief review of the thermodynamics of non-rotating charged BTZ black hole. The effect of thermal fluctuations on various equations of states of charged BTZ black hole is
 seen via the derivations of
  various thermodynamic variables following the same trend as that of uncharged and rotating one.
  The effect of quantum fluctuations, on the entropy, is similar
  to that as on the entropy of uncharged rotating BTZ black hole. For free energy of charged BTZ black hole
  ,  two critical points  are seen to exist. The first one occurs on the horizon axis while the other one occurs
  in the negative region.
Before the first critical point, opposite to the negative correction parameter, the free energy with negative correction parameter,
takes
negative asymptotic value. After the first
critical point,  the correction parameter does not play a significant difference in the free energy.
    Perturbed enthalpy for charged BTZ black hole is calculated, and it is found that for black holes of the small event horizon,
    the enthalpy takes negative (positive) asymptotic
  value for  the positive (negative) correction parameter.  We
  also observed the existence of a critical point at a small horizon radius. From the perturbed enthalpy, we derived perturbed volume.
  We then
  made the use of corrected enthalpy and volume with pressure as an independent variable, for the
  derivations of other thermodynamic potentials like internal energy and Gibbs free energy. From the perturbed Gibbs free energy,
  we observed a critical point  beyond which Gibbs free energy remained constant. Before, the critical point,
   the correction parameter of positive nature, decreased the Gibbs free energy asymptotically. On the contrary, the negatively valued correction
   parameter  asymptotically increased the Gibbs free energy.  Corrected internal energy is then studied and it was found that

the positive value of the correction parameter  did not affect the internal energy much and  made the least difference by following the
same trend as that of
uncorrected internal energy curve. On the other hand, the negative value of the correction parameter yielded a negative asymptotic value
of internal energy.
However as the size of the black hole went  on increasing, the difference between the uncorrected and corrected internal energy got
minimized. Finally, we investigated the stability of charged BTZ black hole and we observed that
 thermal fluctuations, affect the specific heat in the same fashion as that for rotating one.
    It would be interesting to explore the effect of
thermal fluctuations on the $P-V$ criticality of the system by considering the BTZ black holes as a
Van der walls fluid system.  This is matter of future investigation.

\end{document}